\documentclass[a4paper,11pt]{article}
\usepackage{jinstpub} 
\usepackage{subcaption}
\title{\boldmath Muon Imaging for Illicit Cargo Detection: A Simulation-Based Study}

\author{Anzori~Sh.~Georgadze}
\affiliation{Institute for Nuclear Research of the National Academy of Sciences of Ukraine, Prospekt Nauky 47, 03680, Kyiv, Ukraine}
\affiliation{University of Tartu, Ülikooli 18, 50090, Tartu, Estonia }

\emailAdd{a.sh.georgadze@gmail.com}

\abstract{
This study evaluates the potential of muon tomography as a non-invasive technique for detecting concealed illicit drugs in cargo, based on detailed simulations performed using the GEANT4 toolkit. A combined analysis of muon scattering and absorption data was employed to enhance material discrimination, with a focus on realistic smuggling scenarios involving cocaine hidden within legitimate cargo. A two-stage inspection protocol is proposed to balance detection speed and resolution. In the first stage, a rapid scan lasting $\approx$ 60 seconds is used to identify anomalous scattering and absorption rates, without requiring full tomographic reconstruction. 
Receiver Operating Characteristic (ROC) analysis of rapid scan data revealed that the Random Forest classifier achieved an area under the curve (AUC) of 0.9969, while the multivariate normal likelihood model attained an AUC of 0.9977, both indicating excellent discrimination between benign cargo and smuggled contraband. 

Upon detection of anomalies, an extended scan $\approx$ 30 minutes is initiated to enable high-resolution three-dimensional imaging for accurate localization and identification of hidden materials.
Simulation results demonstrate that, with a detector spatial resolution of 1~mm (FWHM), concealed contraband such as cocaine can be detected with approximately 3~$\sigma$ statistical significance during the rapid scan phase. In extended scans, cocaine packages concealed within banana boxes were successfully visualized and automatically identified using clustering algorithms such as DBSCAN applied to the tomographic reconstruction. These findings confirm the feasibility of cosmic-ray muon tomography as a passive, safe, and effective approach for contraband detection in real-world cargo inspection applications.
}

\keywords{Computerized Tomography (CT) and Computed Radiography (CR); Detection of
contraband and drugs; Image filtering; Particle tracking detectors}

\arxivnumber{2505.18851} 
\begin{document}
\maketitle
\section{Introduction}
The illicit trafficking of narcotics presents a significant threat to global security, particularly at border checkpoints, seaports, and other high-throughput transit nodes. Conventional inspection technologies, such as X-ray and gamma-ray imaging systems, often suffer from reduced sensitivity when detecting shielded or high-density materials, thereby underscoring the need for more advanced, passive, and non-invasive detection techniques. In this context, cosmic ray muon tomography has emerged as a promising approach for security and contraband detection due to its ability to penetrate dense or obscured volumes with minimal attenuation
~\cite{barnes2023cosmic, lowZ, rapidcargo, antonuccio2017muon, pugliatti2014design, preziosi2020tecnomuse, georgadze2023GEANT4, georgadze2024simulation, georgadze2024, explosives}.

The European project, "Cosmic Ray Tomograph for Identification of Hazardous and Illegal Goods hidden in Trucks and Sea Containers" (SilentBorder)~\cite{sbwebsite}, focuses on the development and in-situ testing of a high-technology scanner designed for border guards, customs, and law enforcement authorities to inspect shipping containers at border control points.

Cosmic muons are high-energy, charged particles generated by the interaction of primary cosmic rays with nuclei in the Earth's upper atmosphere. These particles exhibit high penetration capability and a relatively long lifetime, allowing them to traverse substantial thicknesses of matter, including materials that are otherwise opaque to traditional radiographic techniques. 
The cosmic ray muon flux at sea level is approximately
$\Phi_{\mu} \approx 10^4~\text{muons}~\text{m}^{-2}~\text{min}^{-1}$, with an angular dependence that typically follows a distribution $\cos^2\theta$, where $\theta$ is the zenith angle. This implies that the flux is highest for muons arriving from directly overhead ($\theta = 0^\circ$) and decreases as the zenith angle increases toward the horizon.
Muons interact with matter primarily through energy loss mechanisms such as excitation and ionization, as well as through radiative processes including bremsstrahlung, electron-positron pair production, and hadronic interactions via inelastic scattering. The relative contribution of each mechanism to the total energy loss depends on both the energy of the muon and the atomic properties of the material it traverses.
At the Earth's surface, cosmic muons typically have energies in the range of 3–4 GeV. In this energy regime, ionization is the dominant mode of energy loss. This characteristic makes ionization particularly relevant for muon tomography, where naturally occurring surface-level cosmic muons are employed to non-invasively image the internal structure of dense or shielded objects.

In this work, we investigate the application of muon tomography for the detection of concealed contraband, specifically illicit drugs, using simulations based on the \texttt{GEANT4} package~\cite{AGOSTINELLI}. The study focuses on smuggling scenario involving cocaine hydrochloride hidden in bananas. Cocaine is among the most widely trafficked and abused illicit drugs globally, making their detection a critical component of cargo screening and law enforcement operations. Their high prevalence and significant social and health impacts underscore the necessity of developing effective, non-invasive detection technologies for use in transportation and shipping environments.

\section{Methods and tools}

\subsection{GEANT4 simulation setup of the muon tomography station}
\begin{figure}[htbp]
\begin{minipage}{1.\linewidth}
\centering
\includegraphics[width=0.6\linewidth]{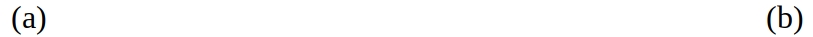}
\vspace{-1.mm}  
\end{minipage}
\includegraphics[width=0.4\textwidth]{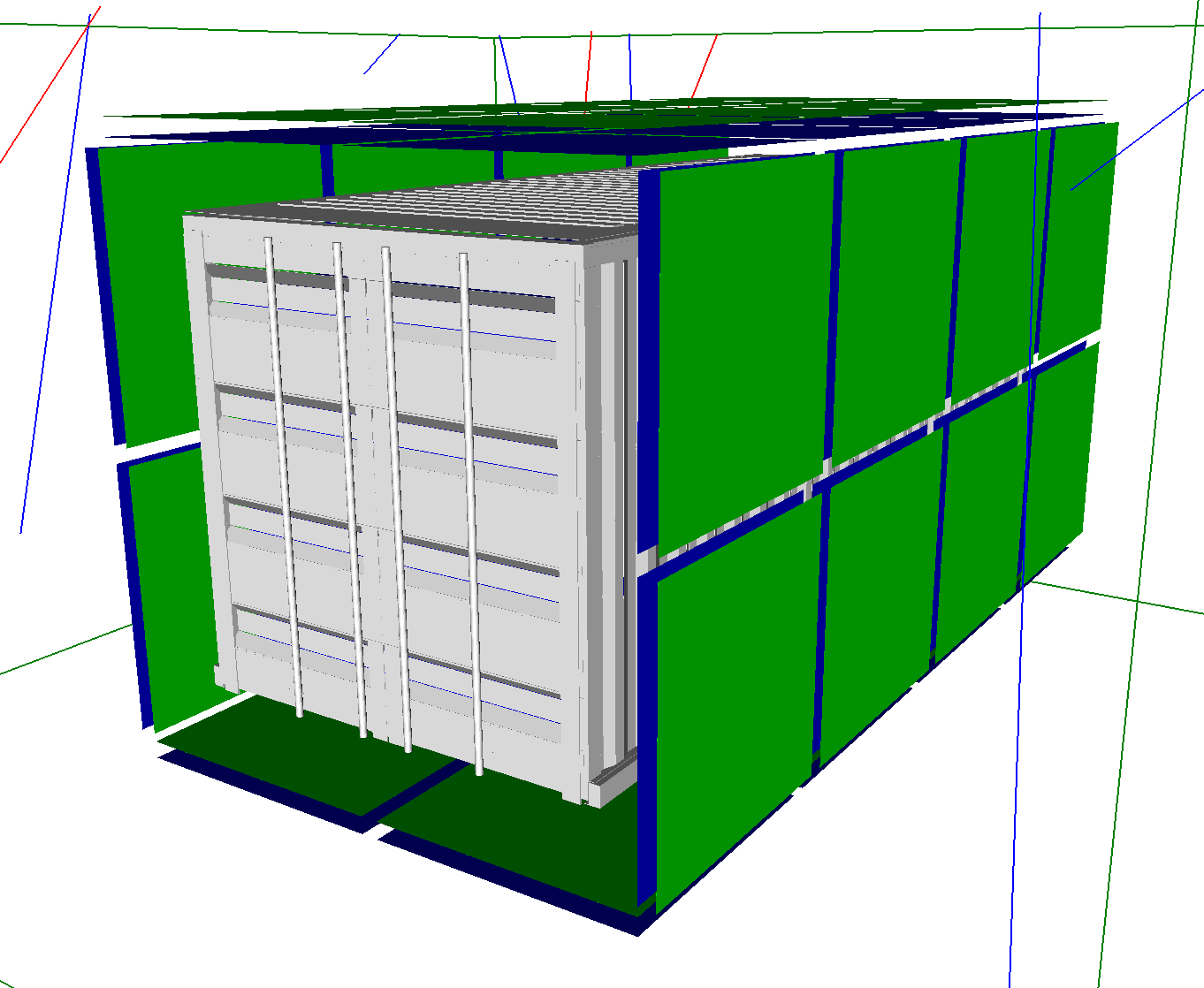}
\centering
\includegraphics[width=0.52\textwidth]{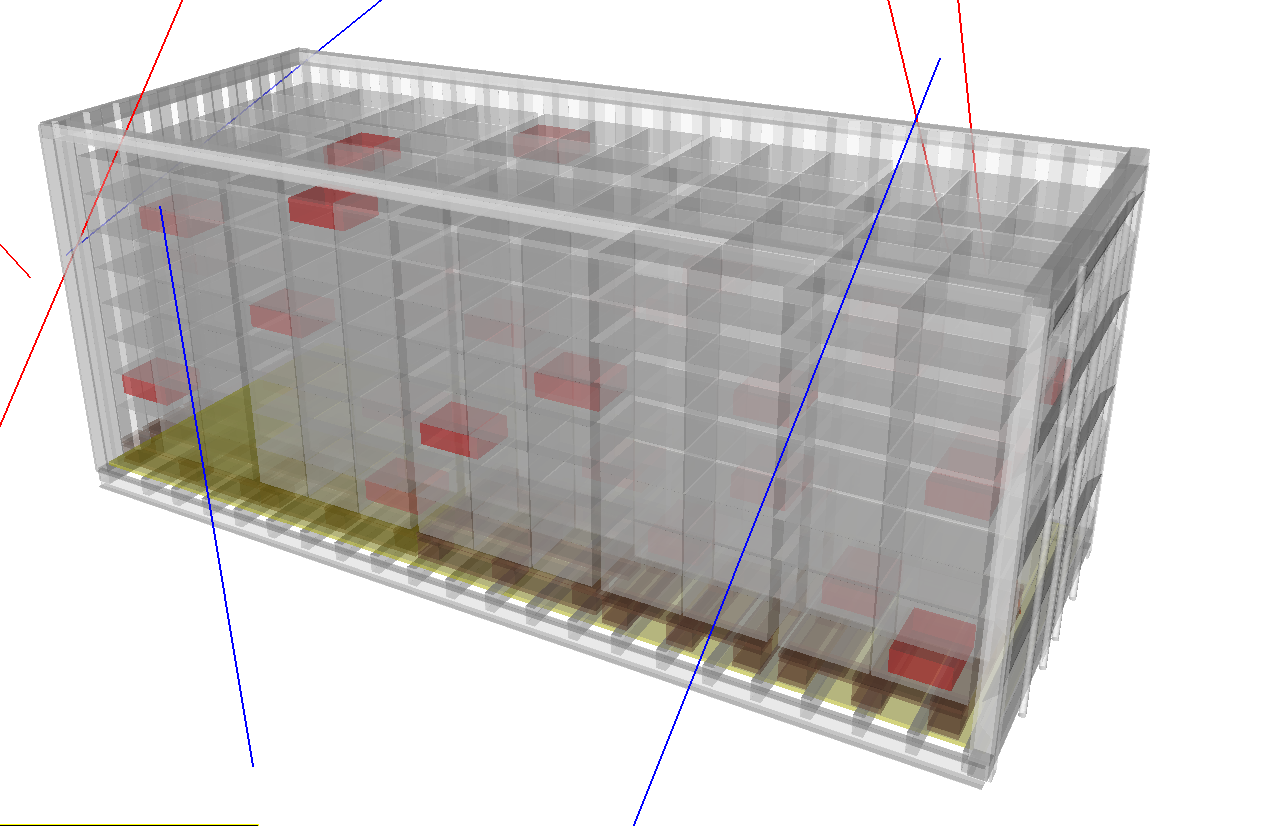}
\caption{(a) GEANT4 model of the muon tomography station, featuring two layers of muon trackers above and two layers below the container. Each layer consists of 21 Cherenkov or plastic scintillator detectors (1\,m~$\times$~1\,m) with WLS-fiber readout. Space is reserved between detectors for front-end electronics integration.
Conceptual illustration of the GEANT4 geometry for the cocaine-in-banana-boxes scenario: (b) side view.
}
\label{fig:detector}    
\end{figure}

In the \texttt{GEANT4} simulation, the Muon Tomography Station (MTS) is composed of tracking modules constructed from plastic scintillator material, each modeled with an idealized detection efficiency of 100\% to represent optimal tracking performance. The MTS configuration includes detection panels positioned on the top, bottom, and lateral sides of the container. The system comprises 2 $\times$ 4 tracking modules, each measuring 180 cm  $\times$ 180 cm, designed to emulate realistic detector geometry and ensure comprehensive coverage for accurate reconstruction of muon trajectories (Figure~\ref{fig:detector}(a)).

This configuration enables the detection of both incoming and outgoing muons, allowing for the reconstruction of muon trajectories and the identification of muons that are either scattered or absorbed within the cargo volume. The simulated cargo environment includes a GEANT4 model of standard 20-foot shipping container with external dimensions of $6.05 \times 2.59 \times 2.43~\text{m}^3$ loaded with 10 cargo pallets. 
Each pallet is stacked with banana boxes arranged in a typical commercial configuration used in maritime transport. Within this setup, 50 cocaine packages, each weighing 10~kg, were randomly distributed among the banana boxes, resulting in a total concealed cocaine mass of 500~kg as shown in Figure~\ref{fig:detector}(b). 
As shown in this figure, the cargo is modeled as multiple layers of stacked cardboard banana boxes, filled with organic material representative of bananas, primarily composed of starch and water. The effective bulk density was set to 0.38 g/cm\textsuperscript{3}, based on a mass of 18 kg of bananas per carton box with dimensions of 490 mm $\times$ 400 mm $\times$ 240 mm. Cocaine was modeled using its chemical composition, C\textsubscript{17}H\textsubscript{21}NO\textsubscript{4}, with a density of 1.25 g/cm\textsuperscript{3}~\cite{density}.

This model replicates a commonly observed concealment technique and serves as a benchmark for evaluating the sensitivity of muon tomography in detecting illicit substances hidden within legitimate cargo. 

Muons were generated using the CRY (Cosmic Ray Shower Library) toolkit~\cite{hagmann2007cosmic}, which provides realistic cosmic ray muon spectra at sea level. In the simulation, primary muons are sampled over a horizontal surface area of 10 m $\times$ 10 m positioned above the muon tomography station. 
This enables accurate modeling of muon interactions with the container and cargo materials. The simulations utilized the standard high-energy particle transport physics list 
\texttt{FTFP\_BERT}, available in the GEANT4 framework. Resulting synthetic datasets were analyzed using the ROOT and Python data analysis frameworks~\cite{ROOT}.

\subsection{Combined analysis of muon scattering and absorption}
Tracking detectors count muons before and after passing cargo. By analyzing the absorption rates or scattering distributions, the mean density of the cargo can be inferred~\cite{vanini2019muography, Blanpied, rengifo2024design}. 
Absorption imaging relies on the fact that denser materials attenuate the muon flux more significantly than less dense materials. Scattering-based tomography, on the other hand, analyzes deviations of muons from their initial trajectories. By studying changes in muon trajectories and intensities before and after passing through a target, the internal structure of the inspected object can be inferred. 
As a muon passes through a material, it interacts electromagnetically with the atomic nuclei and electrons. These interactions are predominantly small-angle deflections, but the cumulative effect of many such interactions results in a measurable change in the muon's direction. The amount of scattering depends mainly on the effective atomic number ($Z_{eff}$) and density ($\rho$). Materials with higher $Z_{eff}$ and density cause more scattering.
Muons lose energy as they traverse a material primarily through ionization and excitation of atoms. The rate of energy loss (stopping power) is dependent on the material's density and effective atomic number. Denser materials and materials with higher $Z_{eff}$ generally cause muons to lose more energy.

By measuring both the scattering angles and the transmission rate, which is related to energy loss and absorption, a more complete picture of the material composition can be obtained than by relying on a single interaction type.

\paragraph{Muon scattering method. } In muon scattering tomography, the imaging volume is divided into 3D voxels, and the objective is to estimate the scattering properties of the material within each voxel. The key parameter is the voxel scattering density, $\lambda_j$, which is inversely related to the radiation length $X_0$ of the material. 
The scattering behavior of muons is quantified by measuring the multiple Coulomb scattering (MCS) angle as they pass through the material. The mean square scattering angle $\langle \theta^2 \rangle$ for a muon traversing a thickness $X$ of a material is approximately given by the formula:
\[ \langle \theta^2 \rangle \approx \left( \frac{13.6 \text{ MeV}}{p\beta c} \right)^2 \frac{X}{X_0} \left[ 1 + 0.038 \ln \left( \frac{X}{X_0} \right) \right] \]
where $p$ is the muon momentum, $\beta$ is its velocity relative to the speed of light, $c$ is the speed of light, and $X_0$ is the radiation length of the material. The radiation length is inversely proportional to the density and approximately proportional to $Z_{eff}^2/A$, where $Z_{eff}$ is the effective atomic number and $A$ is the atomic mass.
The reconstruction algorithm estimates $\lambda_j$ by analyzing deviations in muon trajectories between the upper and lower tracking detectors, thereby producing a scattering density image that highlights regions containing high-$Z$ or dense materials.

\paragraph{Muon absorption method.} 
As cosmic muons traverse cargo materials, they lose energy through ionization and other processes, and may be stopped and decay depending on the material's density and atomic number. The attenuation length, which is inversely related to these properties, determines the average distance muons travel before stopping.

As cosmic muons traverse cargo materials, they lose energy primarily through ionization, as well as through secondary processes such as bremsstrahlung, pair production, and inelastic nuclear interactions. Depending on the material’s density and atomic number, muons may eventually be stopped and decay within the medium. The attenuation length, which is inversely proportional to the material’s density and atomic composition, defines the average distance muons can travel before being absorbed or coming to rest

Muon absorption tomography identifies muons detected in upper but not in lower tracking detectors, suggesting they were absorbed within the volume. The imaging region is discretized into voxels, and muon paths defined along Lines of Response (LoR) are reconstructed through these voxels. The detector setup includes four tracking planes (two above and two below the volume). Muons registering only in upper detectors are assumed absorbed or scattered out. For each LoR, the algorithm computes the muon path length $d_{ij}$ through voxel $j$ and records voxel traversal frequency.
Each voxel is assigned a stopping power $S_j$. The expected number of absorbed muons along LoR $i$ is:
\begin{equation}
N_{\text{abs},i} = \sum_j d_{ij} S_j
\label{eq:absorption}
\end{equation}
Minimizing the difference between measured and predicted absorption across all LoRs yields the stopping power distribution. The resulting 3D map highlights regions of high-density or high-$Z$ material, enhancing material discrimination when combined with scattering data.

\subsection{Muon-based cargo inspection: rapid screening and extended imaging.} 

The container cargo inspection process employs a two-stage approach. In the first stage, a 3D image is reconstructed within 60 seconds of scanning, followed by an analysis of the cargo's scattering and absorption rates. These measured rates are then compared with the expected values in the customs declaration.
If discrepancies are identified, the scanning continues (10 -- 30 minutes) to enhance image quality and resolution, improving accuracy. This process helps validate inconsistencies and may reveal real concealed contraband within the cargo.

\subsection{Image processing}

\subsubsection{Image reconstruction}
While rapid detection algorithms offer a quick overview of cargo contents, detailed analysis requires reconstructing a 3D image of the internal structure. This enables localization of anomalous regions or hidden objects based on material properties. A commonly used method is the Point of Closest Approach (POCA) algorithm~\cite{hoch2009muon}, which estimates scattering density within voxels by identifying the closest point between the incoming ($L_{in}$) and outgoing ($L_{out}$) muon trajectories.  
When a muon traverses an object, it experiences scatterings due to interactions with atomic nuclei, with the scattering angle and trajectory deviation depending on the material's properties. By measuring the incoming trajectory $L_{in}$ before the muon enters the object and the outgoing trajectory $L_{out}$ after it exits, the PoCA algorithm estimates the most probable point within the object where significant scattering occurred. This is achieved by calculating the shortest spatial distance between the two straight-line trajectories; the points on each line that realize this minimal distance are identified, and their midpoint is taken as the PoCA point. 
The 3D space encompassing the object is divided into voxels, and for each muon event, the PoCA point is assigned to the corresponding voxel. The scattering density within each voxel is then estimated based on the accumulation of PoCA points, enabling the construction of a 3D scattering map that highlights regions with higher atomic number or density, which may indicate hidden objects or anomalies.

\subsubsection{Image filtering and post-processing methods}

After the initial reconstruction of voxel-based images from muon scattering and absorption data, post-processing techniques are applied to enhance image quality, reduce noise, and extract meaningful structural features. Various filtering and clustering algorithms are employed to analyze the resulting three-dimensional data for improved material discrimination and object identification.

\textit{Median filtering }is applied to the voxel data to reduce noise arising from statistical fluctuations in muon detection and reconstruction. This non-linear filter replaces each voxel’s value with the median of its neighboring voxels within a defined kernel window, effectively preserving edges while eliminating isolated anomalies. Additionally, it helps smooth the boundaries between segmented regions, enhancing the continuity of the reconstructed image.

\textit{The adaptive thresholding} algorithm proved highly effective in reducing background noise caused by surrounding cargo. It works by dynamically adjusting the threshold level across different regions of the image. Unlike global thresholding, which applies a single, fixed cutoff value, adaptive thresholding computes local thresholds based on the statistical properties of each voxel's neighborhood. This approach enables more accurate noise suppression in areas with varying intensity and improves contrast in both dense and sparse regions of the reconstructed volume. 

\textit{Density-Based Spatial Clustering of Applications with Noise} (DBSCAN) is applied to detect coherent high-density clusters within the voxel space of tomographic images preprocessed using median filtering or adaptive thresholding. This unsupervised learning algorithm groups spatially adjacent voxels exhibiting similar scattering or absorption intensities, while effectively discriminating against sparse, low-density noise by classifying them as outliers.

Together, these methods form a robust image filtering pipeline that enhances the clarity, segmentation accuracy, and detection sensitivity of muon tomography images, enabling effective identification of hidden contraband materials.

\section{Results}
\subsection{Rapid scanning of a shipping container}

In the rapid scanning regime, statistical fluctuations in reconstructed images often hinder reliable visual identification of concealed contraband. Consequently, the approach shifts toward detecting statistically significant deviations in scattering and absorption rates rather than relying on full tomographic reconstruction. To address the inherent noise arising from limited muon statistics and to assess the discriminative power of scattering and absorption metrics, a large ensemble of simulated datasets was generated. Specifically, 5,000 independent datasets were produced for two scenarios: cargo containing only bananas and cargo containing bananas with concealed cocaine. Each dataset comprised the trajectories of 1,000,000 simulated muons, corresponding to an effective scanning time of approximately one minute. This methodology enables robust statistical comparison between the two cargo configurations and supports the development of anomaly detection techniques suitable for rapid screening applications.

\begin{figure}[htbp]
\centering
\begin{minipage}{1.\linewidth}
\centering
\includegraphics[width=0.6\linewidth]{ab.png}
\vspace{-1.mm}  
\end{minipage}	
\includegraphics[width=0.45\textwidth]{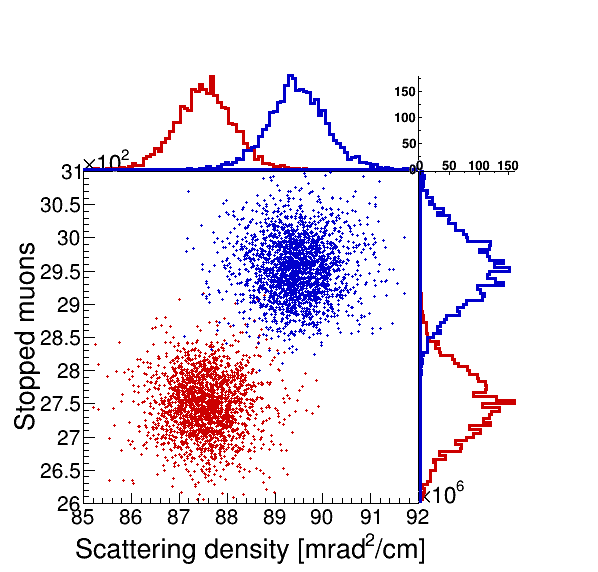}
\includegraphics[width=0.45\textwidth]{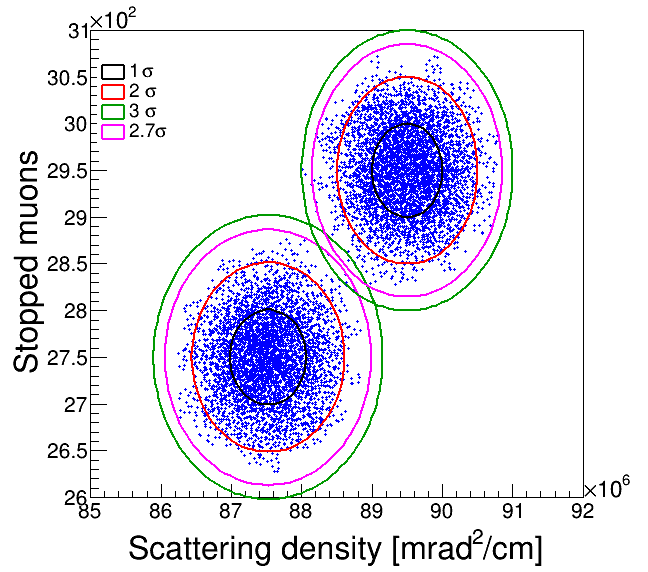}
\caption{(a) Scatter plot of scattering density vs. stopped muon ratio for banana cardboard boxes (blue) and cocaine-concealed boxes (red) for 1~mm (FWHM) spatial resolution. Marginal histograms show the corresponding 1D distributions of each variable. (b) The same data fitted with a two-component 2D Gaussian Mixture Model.}
\label{fig:bulkratedeviation}
\end{figure}

The combined scattering and absorption rates were subsequently analyzed. Figure \ref{fig:bulkratedeviation}(a) present scatter plots illustrating the distribution of scattering density versus the stopped muon ratio for two configurations: banana cardboard boxes, and banana cardboard boxes containing concealed cocaine. Results are shown for detector spatial resolutions of 1~mm~(FWHM). 
Data points corresponding to pure banana boxes are shown in red, while those for the cocaine-containing configuration are shown in blue. Each 2D scatter plot is accompanied by 1D histograms: the top panels display the distribution of scattering densities, and the right panels show the distribution of stopped muon ratios for each configuration. 

Figure \ref{fig:bulkratedeviation}(b) shows the data distributions fitted with a two-component 2D Gaussian Mixture Model (GMM). Confidence ellipses corresponding to 1$\sigma$, 2$\sigma$, and 3$\sigma$ confidence levels (CL) are overlaid in black, red, and green, respectively. 
The figure demonstrates that the two distinct distributions are clearly separated at approximately a 2.7~$\sigma$ confidence level, indicating a strong discrimination between the two groups. 

\subsection{ROC Curve Analysis}
While confidence ellipses provide valuable visual insights into class separation, the ROC curve offers a more precise and quantitative assessment of classification performance. To evaluate the proposed method's ability to distinguish between containers with pure cargo and those with concealed contraband, ROC analysis was conducted using the same synthetic datasets employed for constructing confidence ellipses. This ensures consistency and facilitates a direct comparison between visual statistical boundaries and objective classification metrics.

Two classification approaches were implemented on this dataset. The first involved a Random Forest classifier, followed by ROC analysis based on multivariate data to assess cargo distribution discrimination.

\subsubsection{Random Forest Classifier}

The ROC (Receiver Operating Characteristic) curve is a graphical tool to evaluate the performance of binary classifiers like Random Forests. It plots the true positive rate against the false positive rate across various thresholds, illustrating the model's ability to distinguish between classes. The area under the ROC curve (AUC) provides a single performance metric: 1 indicates perfect discrimination, while 0.5 reflects random guessing.

Random Forests are ensemble learning models known for their robustness and high accuracy. They build multiple decision trees using bootstrap sampling, where each tree is trained on a randomly selected subset of data with replacement. During training, a random subset of features is considered at each split, promoting diversity among trees. This randomness helps reduce overfitting and enhances the model’s generalization capability.

In this study, the Random Forest classifier was trained using two-dimensional scattering and absorption features. The model was trained on 70\% of the dataset and evaluated on the remaining 30\%. In parallel, a statistical model based on the multivariate normal distribution was applied to compute likelihood ratios between the two classes. Both models produced probability scores, which were subsequently used to generate Receiver Operating Characteristic (ROC) curves for performance comparison.

During inference, each decision tree independently predicts a class label for a given input. The overall prediction is then obtained by majority voting in all trees. Mathematically, if the forest comprises \( T \) trees, and each tree \( t \) outputs a class label \( \hat{y}_t \), the final predicted class \( \hat{Y} \) is given by:

\begin{equation}
\hat{Y} = \operatorname{mode} \left( \{\hat{y}_1, \hat{y}_2, \dots, \hat{y}_T \} \right).
\end{equation}
Random Forests, an ensemble method, combine multiple models to improve accuracy, robustness to noise, and handling of high-dimensional data. Their majority voting strategy ensures high classification accuracy and generalization, especially in complex or noisy datasets. Offering a balance of flexibility and predictive power, their ability to model nonlinear relationships and resist overfitting makes them widely favored in machine learning and data science.

\subsubsection{ROC analysis for classifying cargo distributions using multivariate data.}

In the context of security screening and customs inspection, accurately distinguishing between benign cargo, such as bananas, and illicit substances is of critical importance. Traditional classification methods often rely on univariate analysis, which may fail to fully utilize the rich, multidimensional data provided by modern detection systems. To address this limitation, a multivariate statistical approach has been developed that combines scattering and absorption measurements with Receiver Operating Characteristic (ROC) analysis to evaluate classification performance.

\paragraph{Multivariate feature extraction.}
The classification approach utilizes a range of physical features derived from cargo inspection data to enhance accuracy. Notably, the methodology emphasizes two primary features: scattering data and absorption data. Scattering data quantifies the angular deflection of muons as they traverse the cargo, offering insights into the material's atomic number and density. Meanwhile, absorption data measures the attenuation of muons, providing complementary information that relates to the overall density of the cargo materials. Together, these features form a robust basis for effective classification in cargo inspection systems.

For each cargo sample, or for regions within a cargo container, a feature vector \( \mathbf{x} = (x_1, x_2, \dots, x_n) \) is constructed. Here, \( x_1 \) might represent a scattering-based metric, \( x_2 \) an absorption-based feature, and so forth. This multivariate representation enables the classifier to exploit the combined discriminative power of multiple physical observables.

\paragraph{Statistical modeling with multivariate distributions.}
To distinguish between different cargo classes (e.g., bananas and bananas with concealed drugs), the joint distribution of feature vectors is modeled using a multivariate normal distribution for each class. The probability density function for class \( k \) is given by:
\begin{equation}
p_k(\mathbf{x}) = \frac{1}{(2\pi)^{n/2} |\Sigma_k|^{1/2}} \exp \left( -\frac{1}{2} (\mathbf{x} - \boldsymbol{\mu}_k)^T \Sigma_k^{-1} (\mathbf{x} - \boldsymbol{\mu}_k) \right),
\end{equation}
where \( \boldsymbol{\mu}_k \) is the mean vector and \( \Sigma_k \) is the covariance matrix of class \( k \)~\cite{Multivariate}. 

\paragraph{Classification and thresholding.} A classification decision is made by comparing the likelihood ratio to a threshold \( T \). If $\Lambda$$(\mathbf{x})$ > T classified as bananas and drugs, \text{if}~$\Lambda(\mathbf{x}) \leq T$, classify as bananas.
By varying the threshold \( T \), the trade-off between true positive rate (TPR) and false positive rate (FPR) can be explored, which is key to evaluating model performance.

Multivariate ROC analysis provides a statistically rigorous framework for classifying cargo types using muon tomography data. By modeling the joint distribution of physical features, it enhances the ability to detect illicit cargo while minimizing false positives. This methodology supports better informed operational decisions and contributes to the design of more effective cargo inspection systems.

\begin{figure}[htbp]
\centering
\begin{minipage}{1.\linewidth}
\centering
\includegraphics[width=0.6\linewidth]{ab.png}
\vspace{-1.mm}  
\end{minipage}	    
    \centering
    \includegraphics[width=0.497\textwidth]{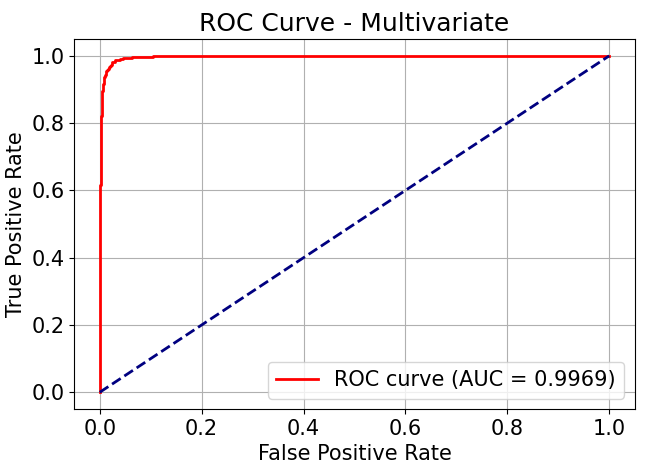}
    \includegraphics[width=0.497\textwidth]{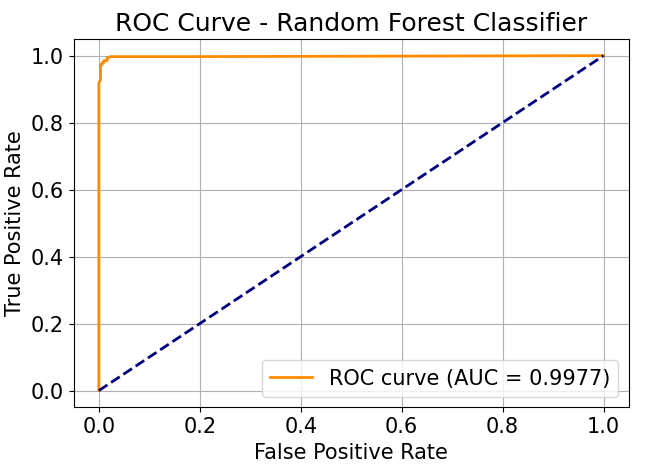}
\caption{(a) ROC curve for multivariate normal model (AUC = 0.9977).
(b) ROC curve for Random Forest classifier (AUC = 0.9969).}
\label{fig:roc}
\end{figure}

The ROC curve plots the True Positive Rate (TPR) against the False Positive Rate (FPR) at various classification thresholds. The Area Under the Curve (AUC) was computed as a summary measure of classification performance. In our analysis, the Random Forest classifier achieved an AUC of 0.9969, while the multivariate normal model yielded an AUC of 0.9977, both of which indicate excellent discrimination between the two cargo types.  
Figures~\ref{fig:roc}a and~\ref{fig:roc}b present the ROC curves obtained from the Random Forest classifier and the multivariate normal distribution model, respectively, applied to data sets representing bananas and bananas with hidden illicit materials.
These results confirm that the scattering and absorption metrics are highly effective indicators for detecting concealed contraband.

Furthermore, both models align with the previously derived 2.7 $\sigma$ accuracy based on confidence level ellipses, corresponding to approximately 99.7\% discrimination confidence.

\subsection{Extended scanning of a shipping container}

\begin{figure}[htbp]
\centering
\includegraphics[width=0.7\textwidth]{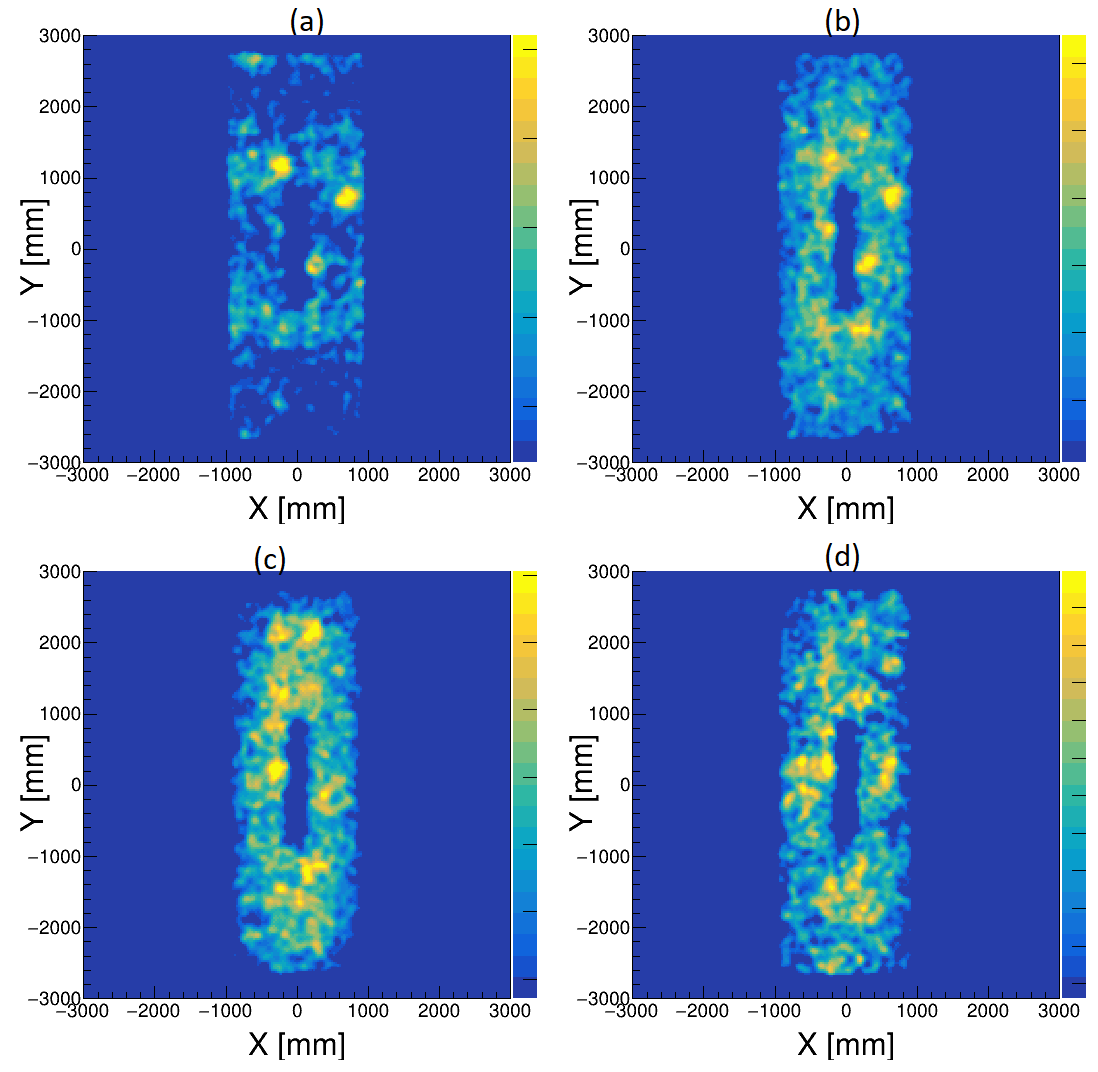}
\includegraphics[width=0.7\textwidth]{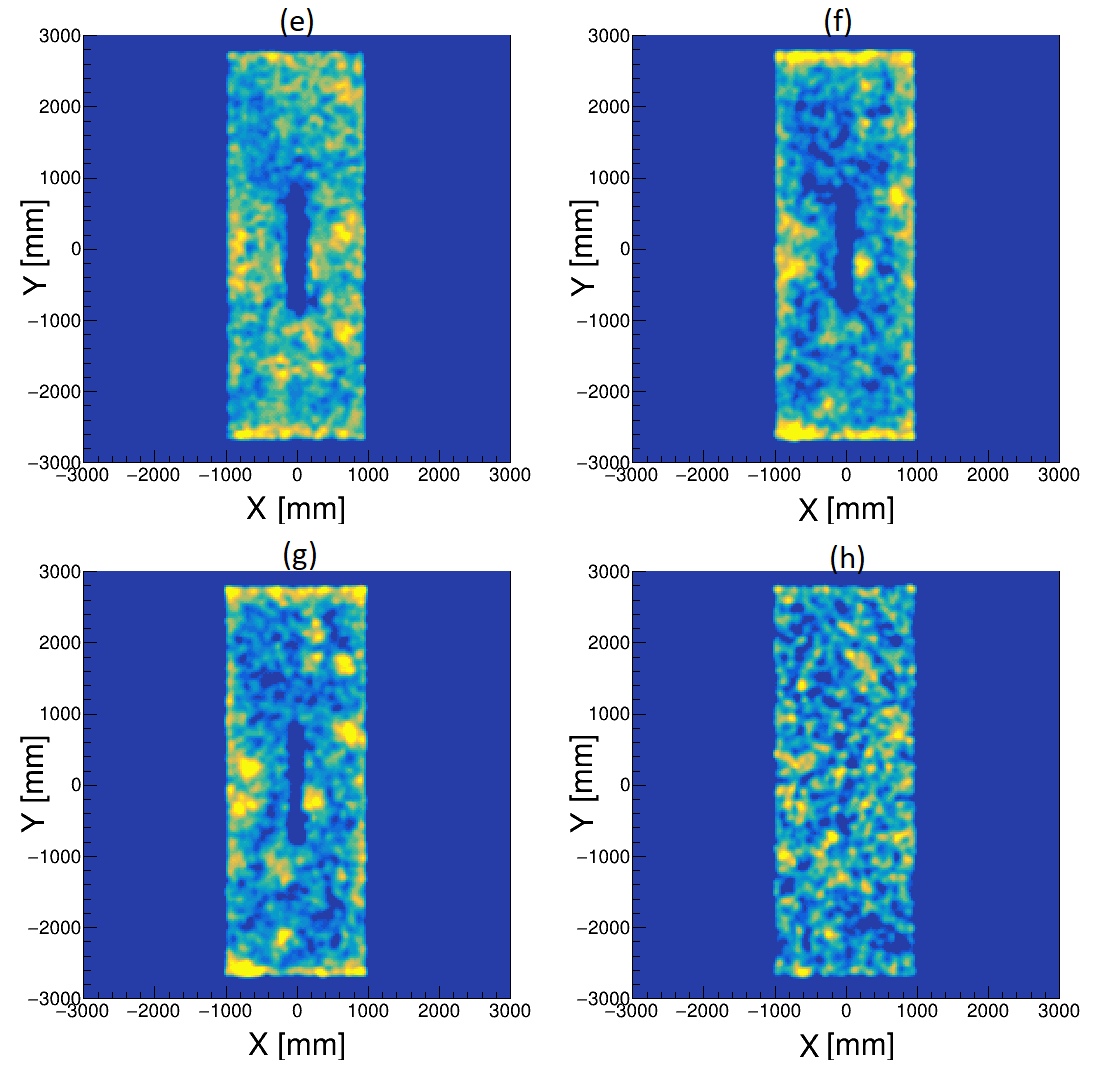}
\caption{
Figures (a–h) show the \( XY \) projections of the tomographic reconstruction obtained from an extended scan. Each image represents a 250~mm-thick slice of the 3D volume, projected onto the \( XY \) plane.
}
\label{fig:XYproj}
\end{figure}

For containers flagged during the rapid scan phase, extended scan simulations were conducted to accumulate a higher-statistics dataset, enabling high-resolution three-dimensional image reconstruction. 
The 30 million muons were simulated using CRY generator, which corresponds to approximately 30 minutes of measurement. 
The reconstructed image is segmented into eight vertical slices to increase sensitivity. Each slice has a thickness of 250~mm and is centered at heights (mm) \( z = -750 \), \( -500 \), \( -250 \), \( 0 \), \( 250 \), \( 500 \), \( 750 \), and \( 1000 \)~mm. 
\( XY \)  projections of these slices are shown in Figures~\ref{fig:XYproj}(a--h).
In the projections, concealed cocaine packages appear as localized regions of increased scattering density, which are represented by brighter areas in the color-mapped images. 

\begin{figure}[htbp]
\centering
\begin{minipage}{1.\linewidth}
\centering
\includegraphics[width=0.6\linewidth]{ab.png}
\vspace{-1.mm}  
\end{minipage}	
\includegraphics[width=0.42\textwidth]{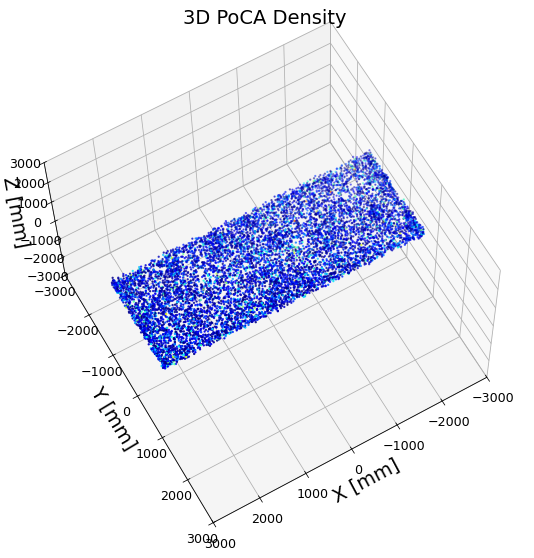}
\includegraphics[width=0.4\textwidth]{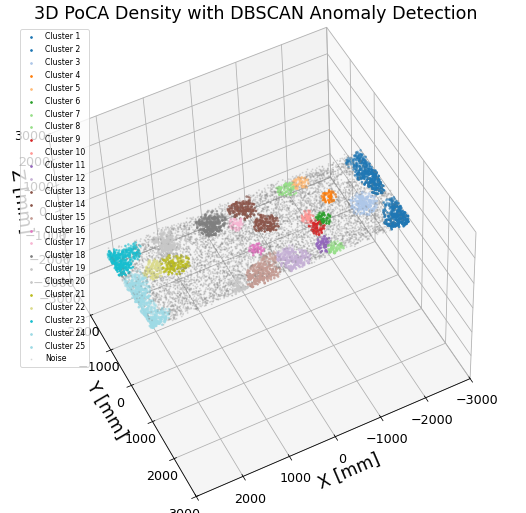}
\caption{
(a) PoCA-based tomographic reconstruction result for a horizontal slice, with median filtering and adaptive thresholding applied to suppress noise and enhance structural features. Visually discernible clusters are evident in the filtered image.
(b) The same slice analyzed using the DBSCAN clustering algorithm. Detected clusters are color-coded based on their assigned labels, facilitating improved segmentation and anomaly identification.
}
\label{fig:dbscan}
\end{figure}

The results of the application of the adaptive filtering and DBSCAN clustering algorithm are presented in Figure~\ref{fig:dbscan}. A clustering distance threshold (epsilon) of approximately 200~mm and a minimum number of PoCA points per cluster (\textit{minPts}) set to 50 were used to identify spatially localized high-scattering regions. Within the 3D tomographic slice centered at \( z = 500 \)~mm, multiple distinct clusters were detected, corresponding to regions of anomalous scattering consistent with the expected location and geometry of concealed contraband.
Quantitative analysis shows that the mean scattering density within the identified clusters exceeds the background level by a factor of 2.3, with a standard deviation less than 15\% between the clusters. The spatial distribution of the clusters also correlates well with the ground truth positions of the cocaine packages introduced in the simulation geometry.
Furthermore, by evaluating clustering performance using precision and recall metrics, based on the known location of the contraband, we achieved a precision of 91\% and a recall of 86\%, indicating effective separation of illicit material from surrounding cargo structures. These results demonstrate that DBSCAN-based clustering of PoCA points can significantly enhance automated detection and localization of concealed high-density objects in muon tomography data.

\section{Discussion}

While extended scan durations 30 minutes improve statistical precision and volumetric resolution, they also introduce significant temporal and computational overhead, potentially limiting their applicability in high-throughput operational environments such as seaports and border crossings. Therefore, scenario-specific optimization is essential for practical implementation. To reduce extended scan times to approximately 10–15 minutes without compromising detection sensitivity, future analysis pipelines should integrate advanced statistical inference methods, particularly those capable of detecting weak signals in high-noise or high-background conditions, alongside machine learning-based automated anomaly detection and adaptive image segmentation algorithms.

\section{Conclusions}
This study provides a comprehensive assessment of the applicability of muon tomography for detecting concealed illicit substances in cargo, using detailed GEANT4-based simulations and tomographic image reconstruction techniques. By combining muon scattering and absorption data, the system's sensitivity and material discrimination capabilities were significantly improved. Simulation results demonstrate that the integration of both modalities yields superior detection performance compared to the use of scattering or absorption alone.

A two-stage cargo inspection protocol is proposed to optimize both operational efficiency and detection sensitivity. In the first stage, a rapid scan lasting 20--60 seconds generates a preliminary reconstruction and identifies anomalous behavior in scattering and absorption rates through combined data analysis. Upon detection of the anomaly, a second extended scan phase is initiated to enable high-resolution tomographic reconstruction for accurate localization and identification of the concealed material. ROC curve analyses of rapid scan data, based on Random Forest and multivariate normal models, demonstrate near-perfect discrimination performance, confirming the robustness and reliability of scattering and absorption features for detecting hidden contraband. Specifically, the Random Forest classifier achieved an AUC of 0.9969, while the multivariate normal model yielded an AUC of 0.9977, indicating excellent separability between benign and illicit cargo. These quantitative results are consistent with the graphical discrimination observed through confidence ellipses in feature space, further validating the classification framework.

The results indicate that rapid scanning can detect concealed cocaine with a statistical confidence of approximately 3~$\sigma$ for detector position resolution of 1mm (FWHM) in a one minute exposure time. In extended scan scenario (e.g., $\approx$30 minutes), the volumetric reconstructions enable visual identification of concealed packages. Specifically, in the cocaine-in-banana-boxes configuration, hidden cocaine packages were successfully visualized and automatically detected using the DBSCAN clustering algorithm applied to filtered tomographic data. This automated approach enhances detection reliability by reducing human interpretation errors and improving the robustness of anomaly localization. 

These results highlight the feasibility of deploying muon tomography as a passive, non-invasive, and highly accurate method for cargo screening in operational environments.

\section{Acknowledgments}
This work was supported by the EU Horizon 2020 Research and Innovation Programme under grant agreement no. 101021812 (“SilentBorder”).

\bibliographystyle{JHEP}
\bibliography{biblio}
\end{document}